\documentclass[12pt,a4paper]{article}
\usepackage[latin1]{inputenc}
\usepackage{amsmath,amssymb,srcltx}
\usepackage[usenames]{color}
\usepackage{amsmath,amssymb,amsthm,amstext} 
\usepackage{hyperref}
\usepackage{graphicx}
\usepackage{caption}

\newcommand{\bt}{\begin{thr}{\bf Theorem. }} 
\newcommand{\satz}{\begin{thr}{\bf Theorem. }\rm}

\theoremstyle{definition}

\makeatletter
\g@addto@macro\@floatboxreset\centering
\makeatother

\begin{document}

\title{Some Statistics concerning the Austrian Presidential Election 2016}

\author{Erich Neuwirth\footnote{Fakult\"at f\"ur Informatik, Universit\"at Wien, W\"ahringer Stra{\ss}e 29, A-1090 Wien, Austria, {\tt erich.neuwirth@univie.ac.at}}
\hspace{20pt}Walter Schachermayer\footnote{Fakult\"at f\"ur Mathematik, Universit\"at Wien, Oskar-Morgenstern-Platz 1, A-1090 Wien, Austria, {\tt walter.schachermayer@univie.ac.at} and the Institute for Theoretical Studies, ETH Zurich. Partially supported by the Austrian Science Fund (FWF) under grant P25815, the Vienna Science and Technology Fund (WWTF) under grant MA09-003 and Dr. Max R\"ossler, the Walter Haefner Foundation and the ETH Zurich Foundation.}}

\date{\today}
\maketitle

\begin{abstract}
\noindent

The 2016 Austrian presidential runoff election has been repealed by the Austrian constitutional court. The results of the counted votes had yielded a victory of Alexander van der Bellen by a margin of 30.863 votes as compared to the votes for Norbert Hofer. However, the constitutional court found that 77.769 votes were ``contaminated'' as there have been - at least on a formal level - violations of the legal procedure when counting those votes. For example, the envelopes were opened prematurely, or not all the members of the electoral board were present during the counting etc. Hence the court considered the scenario that the irregular counting of these votes might have caused a reversal of the result as {\it possible}. The constitutional court sentenced that this {\it possibility} presents a sufficient irregularity in order to order a repetition of the entire election.

While it is, of course, {\it possible} that the irregular counting of those 77.769 votes reversed the result, we shall show that the probability, that this indeed has happened, is ridiculously low.
\end{abstract}

\section{Introduction}

On May 22, 2016 the Austrians voted in a runoff election between Norbert Hofer (candidate 1) and Alexander van der Bellen (candidate 2). The result after counting the votes was 49.7~:~50.3 in favor of van der Bellen. For the precise data we refer to \cite{wahl16}. 

The party supporting the candidate Norbert Hofer subsequently appealed to the Austrian constitutional court, claiming irregularities in the procedure of counting the mail votes.

Here are the details. In Austria there are two ways of casting one's vote. Either by showing up personally at the poll site and delivering the vote into the ballot box (ballot voting), or by sending the vote by mail during a well-defined period preceding the voting day (mail voting).

The allegation of Hofer's party was that in some districts the counting of the mail votes violated the procedure stipulated by the law. For example, the letters of the outer envelopes containing these votes (in an inner envelope) should not be opened before 9:00~a.m.~of the subsequent Monday, May 23. The reason is that the electoral board for the mail votes for districts only is called on duty for this time. By opening these letters prematurely these votes became invalid was argued by the alleging party. Several other accusations were made, involving different degrees of severity \cite{vfgh}.


The constitutional court carefully investigated these accusations and concluded that in 11 of the 117 voting districts there have indeed happened violations of the law during the procedure of counting the mail votes. The court sentenced that in total there were 77.769 mail votes counted in an irregular way. The central argument of the court in favor of ordering a repetition of the election was that there was the {\it possibility} that manipulations on such a number of votes might have led to a reversal of the result. After all, the margin was only 30.863 votes.

On the other hand, the constitutional court states explicitly in its findings that there was no evidence that there actually have been  any manipulations of the votes. What has been proved were several violations of the legally prescribed procedure of counting the votes.

\section{The Analysis}

Our goal is to obtain a quantitative analysis of the probability that there was in reality a victory of Norbert Hofer and was only turned afterwards -- in whatever way -- into a victory of Alexander van der Bellen because of wrong-counting the mail votes in the incriminated 11 districts. 

To do so, we first compare the results in the $N = 106 = 117-11$ ``green'' or ``uncontaminated'' districts where the court did not find violations of the legal procedures, with the $M = 11$ ``red'' or ``contaminated'' districts  districts where the court found violations of these procedures.

\begin{figure}[h]
  \includegraphics{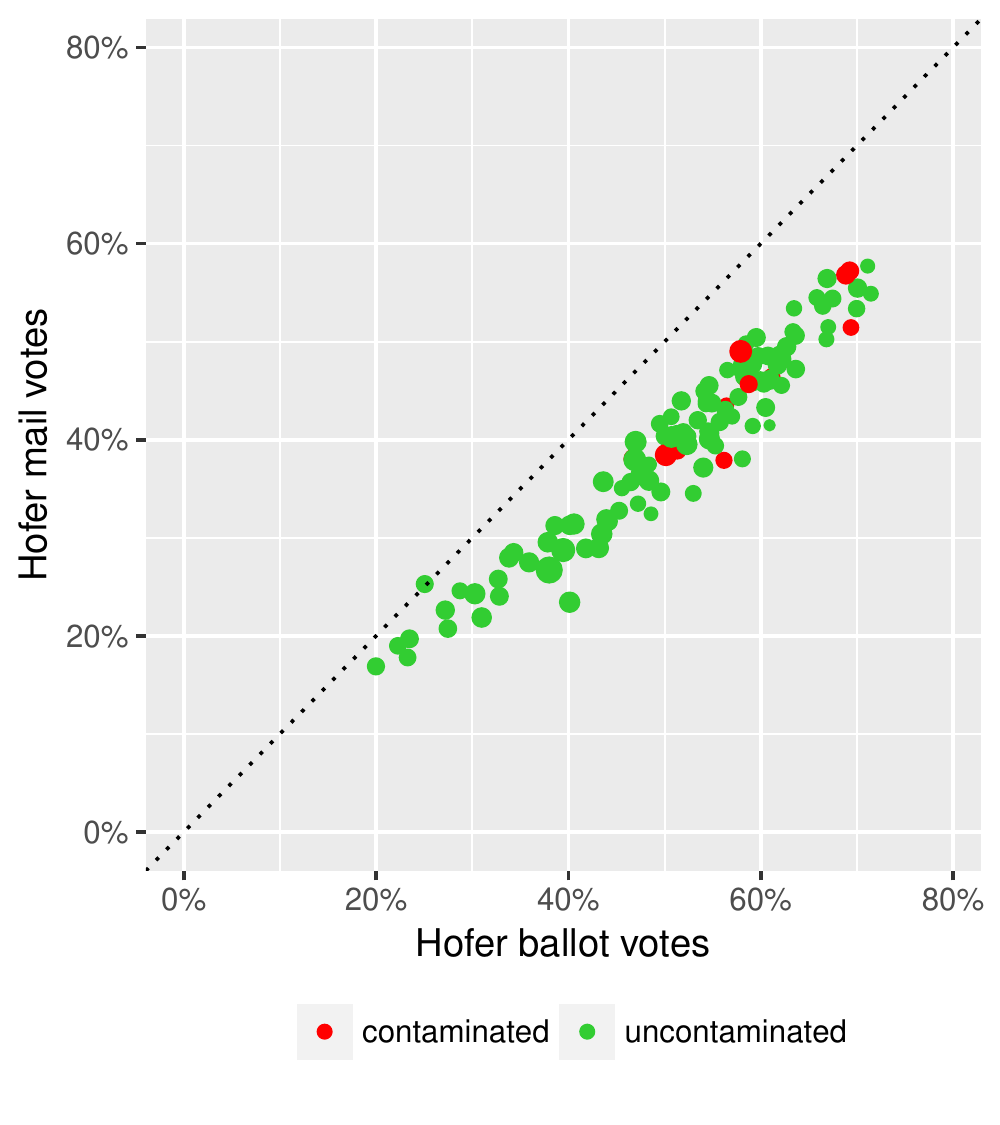}
  \caption{Mail and ballot vote percentages - official results}
\label{figure:diag1}
\end{figure}

Each ``green'' district corresponds to a green dot: on the x-axis we plot the percentage of votes for candidate 1 (Norbert Hofer) among the ballot votes, and on the y-axis the percentage of the votes for candidate 1 among the votes by mail. The picture clearly indicates a linear relation between these two ratios. One also sees that the slope of the regression line is smaller than 1 which does not come as a surprise. Among the voters of Norbert Hofer the propensity to use the possibility of voting by mail is smaller than among the voters of Alexander van der Bellen. We can also observe that the intercept of the regression line essentially vanishes, i.e. the regression line essentially passes through the origin. 

While the green dots correspond to the ``uncontaminated'' districts, the red dots in Figure \ref{figure:diag1} correspond to the 11 ``contaminated'' ones. Glancing at Figure \ref{figure:diag1} one cannot see any alarming behavior of the red dots.

The above Figure \ref{figure:diag1} corresponds to the {\it counted} votes. In the subsequent analysis we shall only accept the green dots as valid data. As regards the red dots, we only take their x-coordinate as granted: recall that the x-coordinate corresponds to the percentage of ballot votes in favor of candidate 1. As regards the y-coordinates of the red dots, it is precisely our point to analyze whether the {\it true} votes gave different results than the {\it counted} votes in a degree which could have resulted in a reversal of the election result.

\begin{figure}[h]
  \includegraphics{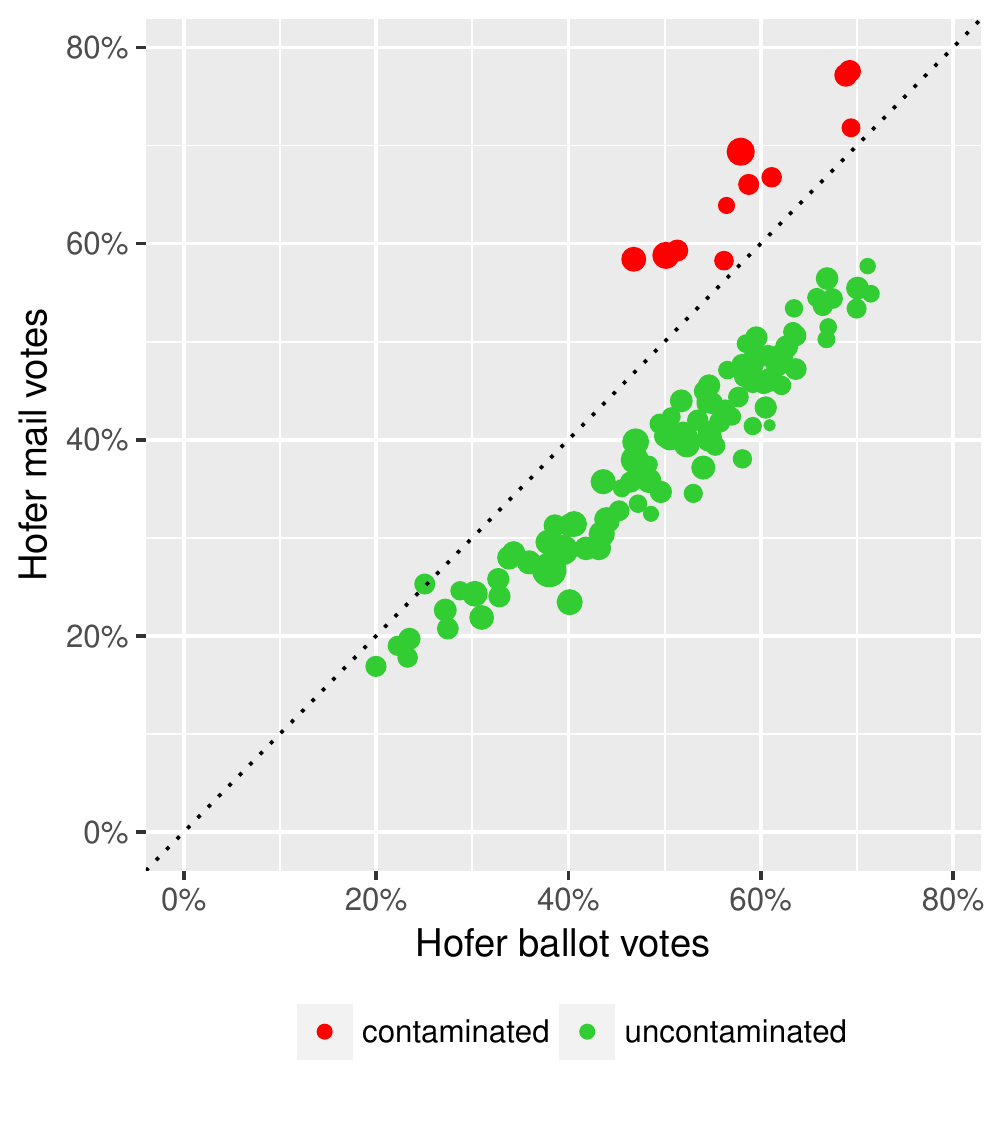}
  \caption{Mail and ballot vote percentages - modified results}
  \label{figure:diag2}
\end{figure}

As an illustration, Figure \ref{figure:diag2} below indicates a scenario for the {\it true} votes which would have yielded a victory for candidate 1 (by a margin of 1 vote).  To obtain Figure \ref{figure:diag2}, we have assigned  -- hypothetically -- 15.432 (half the missing 30.863 votes, rounded up)  proportionally to the 11 ``contaminated'' districts, and subsequently recalculated the corresponding percentages. This procedure  implements a scenario where that many votes have wrongly been
counted for candidate 2 instead of candidate 1.
For more detailed information we refer to the web site of the first named author \cite{EN} (\url{ http://www.wahlanalyse.com/WahlkartenDifferenzenVfGh.html}).

It is evident that a scenario as in diagram 2 does not look very likely to have happened in reality. This diagram only has an illustrative character for our purposes in order to visualize the absurdity of such a scenario. It will not play any role in the subsequent analysis. In particular, we shall not assume a certain given assignment of the missing 30.863 votes to the 11 red districts. We shall only be interested in their {\it total sum}. Speaking mathematically, we shall eventually calculate the probability distribution of the total number of {\it true} mail votes in the incriminated districts, conditionally on the given results of the ``green'' districts, and calculate the probability of the event that they would have resulted in a victory of candidate 1.

To analyze the probability that the {\it true} votes by mail in the red districts would have yielded a victory for candidate 1, we apply a weighted linear regression model to the green dots in diagram 1.

In fact, since the calculations become easier to write and to program, we use a regression model on the number of votes
instead of the percentages, which is mathematically equivalent. Since the expected variation of the votes depends on the total number of votes in the respective districts,
the model exhibits heteroskedascity, and we have to use a weighted regression.

Regression models include prediction intervals for observations with known values for the independent variable(s) and known standard deviations (compared to standard deviations of completely known cases).
Using these procedures and assuming that the mail vote results in the contaminated districts follow the model of the uncontaminated districts, we can then compute the distribution of the sum of the expected
votes in the contaminated 11 districts (which, under the present model assumptions, follows a rescaled
$t$-distribution).

Using the distribution of this random variable, we can then  calculate the probability that the {\it true votes} would have resulted in an election of candidate 1.
As we shall see below, the numerical value equals $p = 1.322065 \cdot 10^{-10}$.

Actually, it turns out that the sentence of the constitutional court may also lead to a slightly different calculation. Apart from the 11 ``contaminated'' districts it identified 3 more ``dubious'' districts where things are not so clear. We refer to \cite{vfgh} for the details. Although the sentence of the court did not take into account these districts, one might argue that -- possibly -- also in these districts the counting of the mail votes was not reliable.  Mathematically speaking, this leads to the consideration of M = 14 ``red'' and N = 117 -14 = 103 ``green'' districts. The calculations then are  identical to the above considered case and lead to a numerical value of $p = 5.151422 \cdot 10^{-8}$.

\section{The Model}

We consider $N=117-11=106$ voting districts with a total of $t_n$ valid votes for $n=1, \dots,N.$ In district $n,$ there were $v_n$ votes counted for candidate 1 and $\bar{v}_n$ votes for candidate 2, so that
$v_n + \bar{v}_n = t_n$. 

The votes $t_n$ split into $b_n$ many ballot votes and $m_n$ many mail votes which again are divided into $v_{b,n} (resp.~v_{m,n})$ many votes for candidate 1 and $\bar{v}_{b,n}=b_n-v_{b,n} \,(resp.~ \bar{v}_{m,n} = m_n - v_{m,n})$ many votes for candidate 2.

Our objects of interest are the vote numbers


$$v_{b,n} \quad \mbox{and} \quad v_{m,n} , \quad n=1, \dots, N.$$


These numbers denote the votes for candidate 1 among the ballot and mail votes respectively. As indicated by the diagrams above, a linear relation between these quantities is justified as model assumption. To make the plots easier to understand, we used percentages instead of votes there.


While the numbers $(v_{b,n})^N_{n=1}$ are considered as given data the numbers $(v_{m,n})^N_{n=1}$ are considered as realizations of the following random variables: 
\begin{equation}\label{M1}
V_{m,n}=k \, v_{b,n} + \epsilon_n, \quad n=1, \dots, N
\end{equation}

Here $k$ is an unknown deterministic number while $(\epsilon_n)^N_{n=1}$ are independent centered Gaussian random variables. Their variance is
${\sigma^2}{m_n}$,
for some (unknown) deterministic number $\sigma>0.$ 

Variances of votes for parties being proportional to the number of total votes is a standard model assumption in statistical voting analysis procedures {(see \cite{bruckmann},
\cite{neuw84}, \cite{neuw94}, \cite{neuw2012}, and \cite{ledl}} ).



We are thus facing a heteroskedastic, linear regression model.

Applying standard regression theory, we obtain the estimators $\hat{k}, \hat{\sigma}$ which we consider as random variables. In particular, the estimator $\hat{k}$ follows a (rescaled) $t$-distribution whose parameters can be explicitly calculated for the given data.

We next consider $M=11$ many ``contaminated'' districts, disjoint from the $N$ `` uncontaminated'' districts.

Assuming that the results $V_{m,j} \quad j=1, \dots, Mm$ also follow model \ref{M1}, they can be considered
as random variables
$$V_{m,j} = k \, v_{b,j}+\epsilon_j, \quad j=1, \dots, M.$$

As we do not know the true value of $k$ we have to consider the estimated variable 
$$\hat{V}_{m,j}=\hat{k}v_{b,j} + \epsilon_j, \quad j=1, \dots, M. $$
The new noise variables $(\epsilon_j)^M_{j=1}$ are such that $((\epsilon_n)^N_{n=1}, (\epsilon_j)^M_{j=1})$ are independent. The variance of the $\epsilon_j$ again is given by $\sigma^2 m_j.$



Finally we consider the sum
$$\hat{V}=\sum^M_{j=1} \hat{V}_{m,j}.$$
$\hat{V}$ is the random variable modeling the total mail votes for candidate 1 in the $M$ ``contaminated'' districts. 

Denoting by $v_b=\sum^m_{j=1}v_{b,j}$ the total number of ballot votes of candidate 1 in the $M$ ``contaminated'' districts, and by $m=\sum^M_{j=1}m_j$ the total number of mail votes in these districts, we may write $\hat{V}$ as 

$$\hat{V}=\hat{k} \, v_b + \epsilon$$
where $\epsilon$ is a centered Gaussian variable with variance $\sigma^2 m$.

A standard tool of regression theory allows us to compute a prediction interval for $\hat{V}.$

If $\sigma^2$ were known, $\hat{V}=\hat{k}v_b + \epsilon$ were normally distributed with mean $\hat{k} v$ and
variance $\sigma^2\left(\frac{v_b^2}{\sum_{n=1}^N{\frac{v_{b,n}^2}{m_n}}}+m\right)$

Applying this fact we use the regression model estimate for $\hat{\sigma}$ as substitute for
the unknown constant $\sigma$. Using this, the random variable


$$\frac{\hat{V}-\hat{k}v_b}{\hat{\sigma}\sqrt{\frac{v_b^2}{\sum_{n=1}^N{\frac{v_{b,n}^2}{m_n}}}+m}}$$
follows a $t$-distribution with 105 degrees of freedom.

We compare this random variable with the critical number $\tilde{V}$ which would be necessary for candidate 1 in order to reverse the result. Finally we compute
\begin{equation*}
\mathbb{P}[V\geq \tilde{V}]
\end{equation*}
which can be computed explicitly and yields the desired result.

\section{Confidence intervals for prediction}

We use results for the standard heteroskedastic linear model

$$y=X\beta+\epsilon$$
with covariance matrix $\textrm(cov)(\epsilon)=\sigma^2 W$ for a given positive definite matrix $W.$

In this model, the best linear unbiased estimator for $\beta$ is
$$\hat{\beta}=(X'W^{-1}X)^{-1}X'W^{-1}y$$

The covariance of this estimator is

\begin{equation*}
\begin{split}
\textrm{cov}(\hat{\beta}) & =(X'W^{-1}X)^{-1}X'W^{-1}\textrm{cov}(y)((X'W^{-1}X)^{-1}X'W^{-1})' \\
& = (X'W^{-1}X)^{-1}X'W^{-1}\sigma^2 W W^{-1}X(X'W^{-1}X)^{-1} \\
& = \sigma^2(X'W^{-1}X)^{-1}
\end{split}
\end{equation*}

In our case, $X$ is the $N$x$1$-matrix $(v_{b,n})_{n=1}^N$ and $W$ is the diagonal matrix
$\textrm{diag}((m_n)_{n=1}^N$). The parameter $\beta$ in our case is the scalar $k$ and its
estimator is $\hat{k}$, and $\textrm{cov}(\hat{\beta})$ becomes $\textrm{var}(\hat{k})$

Therefore $X'W^{-1}X=\sum_{n=1}^N v_{b,n}\frac{1}{m_n}v_{b,n}=\sum_{n=1}^N \frac{v_{b,n}^2}{m_n}$ and
$$ \textrm{var}(\hat{k})=\sigma^2\frac{1}{\sum_{n=1}^N \frac{v_{b,n}^2}{m_n}}$$

We want to compute the distribution of $\hat{V}=\hat{k} v_b + \epsilon$ with $\textrm{E}(\epsilon)=0$ and $\textrm{var}(\epsilon)=\sigma^2 m$.

We have 
\begin{eqnarray*}
\textrm{E}(\hat{k} v_b + \epsilon) & = & \textrm{E}(\hat{k}v_b) + \textrm{E}(\epsilon)  =  k v_b \\
\textrm{var}(\hat{k} v_b + \epsilon) & = & \textrm{var}(\hat{k} v_b) + \textrm{var}(\epsilon) =  \sigma^2\left(\frac{v_b^2}{\sum_{n=1}^N \frac{v_{b,n}^2}{m_n}}+m\right)
\end{eqnarray*}

Since $\sigma^2$ is unknown, we have to replace it by the estimator $\hat{\sigma}^2$ and then $\frac{\hat{V}-\hat{k}v_b}{\hat{\sigma}\sqrt{\frac{v_b^2}{\sum_{n=1}^N \frac{v_{b,n}^2}{m_n}}+m}}$ follows a $t$-distribution with $N-1$ degrees of freedom. From that
confidence intervals for $\hat{V}$ can be derived easily.

\section{The Results}

All the data and the code for performing our analysis can be found at \\
\url{https://github.com/neuwirthe/AustrianPresidentialElection}.

Candidate 1 had $34479$ votes in the contaminated districts, and he would need additional $15432$ votes, so that
in total he needs $\tilde{V}=34479+15432=49911$ votes to overturn the result.

Using the R code from the URL above to compute the probability of
a result overturning the result in favor of Hofer, we get the value

$$p= 1.322065\cdot 10^{-10}. $$


\begin{thebibliography}{99}  

\bibitem{wahl16} 
\url{http://www.bmi.gv.at/cms/BMI_wahlen/bundespraes/bpw_2016/}

\bibitem{EN} 
\url{http://www.wahlanalyse.com/WahlkartenDifferenzenVfGh.html}

\bibitem{bruckmann}
Bruckmann, G. (1966). \textit{Sch\"atzung von Wahlresultaten aus Teilergebnissen} Wien: Physica-Verlag.


\bibitem{ledl}
Ledl, Th. (2007). \textit{Modellierung von Wechselw\"ahlerverhalten als Multinomialexperiment} Dissertation, Fakult\"at f\"ur Wirtschaftswissenschaften und Informatik, Universit\"at Wien. \url{http://homepage.univie.ac.at/thomas.ledl/download/Dissertation.pdf} 

\bibitem{neuw84}
Neuwirth, E. (1984). \textit{Sch\"atzung von W\"ahler\"ubergangswahrscheinlich\-keiten} In: M. Holler (Hg.) Wahlanalyse -- Hypothesen, Methoden und Ergebnisse. M\"unchen: tuduv-Buch.

\bibitem{neuw94}
Neuwirth, E. (1994). \textit{Prognoserechnung am Beispiel der Wahlhochrechnung} In: P. Mertens. (Hg.) Prognoserechnung. W\"urzburg-Wien.


\bibitem{neuw2012}
Neuwirth, E. (2012), \textit{Wahlhochrechnung: ein kurzer \"Uberblick \"uber den Einsatz bei bundesweiten Wahlen in \"Osterreich}, in: \"Osterreich 2032 (Festschrift zum 80. Geburtstag von Gerhart Bruckmann), Hrsg: Lutz, W. und Strasser, H., Verlag der \"osterreichischen Akademie der Wissenschaften, Wien 2012.


\bibitem{vfgh}
  Verfassungsgerichtshof, Freyung 8, A-1010 Wien, \textit{W I 6/2016-125}, \\ 1.~Juli~2016, \\
  \url{https://www.vfgh.gv.at/cms/vfgh-site/attachments/5/7/8/CH0003/CMS1468412977051/w_i_6_2016.pdf}


\end{thebibliography}
\end{document}